\documentclass[%
reprint,
nofootinbib,
 amsmath,amssymb,
 aps,
]{revtex4-2}

\usepackage{graphicx}
\usepackage{dcolumn}
\usepackage{bm}
\usepackage{hyperref}

\usepackage{subfig, tabularx, afterpage,xifthen}
\usepackage{amsmath,amsfonts,amssymb,mathtools}
\usepackage{slashed}

\usepackage{xcolor}

\newcommand\scetone{SCET$_\mathrm{I}$} 
\newcommand\scettwo{SCET$_\mathrm{II}$} 

\newcommand\bra[1]{\left\langle {#1} \right| }
\newcommand\ket[1]{\left| {#1} \right\rangle}

\newcommand\nbar{{\overbar{n}}}
\newcommand\nb{{\overbar{n}}}
\newcommand{\overbar}[1]{\mkern 1.5mu\overline{\mkern-1.5mu#1\mkern-1.5mu}\mkern 1.5mu}

\newcommand\myvec[1]{\textbf{#1}}

\newcommand\sub{\mathrm{sub}}

\newcommand\alphabar{\overbar{\alpha}}

\newcommand\Mmt{\mathcal{M}}

\newcommand\zbar{\overbar{z}_1}
\newcommand\zzbar{\overbar{z}_2}

\newcommand\QuarkWout{W^\dagger}

\newcommand\QuarkWinc{\overline{W}}

\def\ns{\fmslash{n}}
\def\nbs{\fmslash{\nbar}}

\def\OMIT#1{{}}
\def\eqn#1{Eq.\ \eqref{#1}}
\def\lqcd{\Lambda_{\rm QCD}}
\def\shat{\hat{s}}
\def\omegaoneb{\overbar\omega_1}
\def\omegatwob{\overbar\omega_2}

\def\plusfn#1{\left[#1\right]_+}
\def\plusmu#1{\left[#1\right]_+^\mu}
\def\ttwo{T_{(0,0)}}
\def\ctwo{C_{(0,0)}}
\def\dtwo{V_{(0,0)}}
\def\gtwo{\gamma_\nu^{(0,0)}}
\makeatletter
\def\fmslash{\@ifnextchar[{\fmsl@sh}{\fmsl@sh[0mu]}}
\def\fmsl@sh[#1]#2{%
   \mathchoice
     {\@fmsl@sh\displaystyle{#1}{#2}}%
     {\@fmsl@sh\textstyle{#1}{#2}}%
     {\@fmsl@sh\scriptstyle{#1}{#2}}%
     {\@fmsl@sh\scriptscriptstyle{#1}{#2}}}
\def\@fmsl@sh#1#2#3{\m@th\ooalign{$\hfil#1\mkern#2/\hfil$\crcr$#1#3$}}

\begin{document}

\preprint{APS/123-QED}

\title{Rapidity Logarithms in SCET Without Modes}

\author{Matthew Inglis-Whalen}
\email{minglis@physics.utoronto.ca}
\author{Michael Luke}
\email{luke@physics.utoronto.ca}
\author{Aris Spourdalakis}
\email{aspourda@physics.utoronto.ca}

\affiliation{Department of Physics, University of Toronto, Toronto, Ontario, Canada M5S 1A7}%

\date{\today}

\begin{abstract}
We re-examine observables with rapidity divergences in the context of a formulation of Soft-Collinear Effective Theory in which infrared degrees of freedom are not explicitly separated into modes. We consider the Sudakov form factor with a massive vector boson and Drell-Yan production of lepton pairs at small transverse momentum  as demonstrative examples. In this formalism, rapidity divergences introduce a scheme dependence into the effective theory and are associated with large logarithms appearing in the soft matching conditions. This scheme dependence may be used to derive the corresponding rapidity renormalization group equations, and rates naturally factorize into hard, soft and jet contributions without the introduction of explicit modes. 
\end{abstract}

\maketitle


\section{Introduction}
 
Effective Field Theory (EFT) offers an elegant framework for systematically separating the physics at different scales in a given process.  When working with a cutoff $\mu$, physics at high energy scales $\mu_{H}>\mu$ is integrated out of the theory, and its effects on physics at lower energy scales $\mu_S<\mu$ is taken into account with a series of effective operators of increasing dimension whose effects are suppressed by powers of the ratios of the two scales.  One advantage of this approach is that observables depending on multiple scales may be systematically factorized into functions that each depend only on a single energy scale and an arbitrary factorization scale $\mu$.
Each factor may then be evaluated at its natural scale, and using renormalization group evolution (RGE) can be brought under perturbative control at an arbitrary scale $\mu$. In multi-scale processes, the theory is matched at each relevant scale $\mu_i$ to a new effective theory where physics at scales above $\mu_i$ is integrated out, allowing physical quantities to be factorized into multiple terms, each of which depends on a single scale.

Soft-Collinear Effective Theory (SCET) 
 \cite{Bauer:2000ew,Bauer:2000yr,Bauer:2001ct,Bauer:2001yt,Bauer:2002nz,Beneke:2002ph,Beneke:2002ni} 
achieves this factorization in hard scattering processes by explicitly introducing separate fields, or modes, for each relevant scaling of the various momentum components of the field. A typical SCET factorization theorem separates physical processes into hard, collinear, and soft/ultrasoft pieces.  Hard physics (above the cutoff) is incorporated as usual into the matching coefficients of operators in the effective Lagrangian, whereas the factorization of low energy degrees of freedom occurs dynamically in the effective theory: soft, ultrasoft and collinear degrees of freedom are described by distinct fields which decouple at leading power in the SCET Lagrangian.  This allows factorization theorems for many observables to be derived.  Processes factorizing into collinear and ultrasoft modes, such as Deep Inelastic Scattering (DIS) in the $x\to 1$ limit, are referred to as \scetone\ processes, whereas those factorizing into collinear and soft modes, such as Drell-Yan (DY) with $q_T^2\ll q^2$, are referred to as \scettwo\  processes. More complicated processes may require additional modes, and have more complex factorization theorems; some examples are given in  \cite{Bauer:2011plus,Procura:2014cba,Procura2018+,Rothstein2016G,Bauer:2011G}.

In \cite{Goerke:2017ioi} it was proposed that the introduction of separate modes in SCET is not necessary to factorize hard processes in QCD, and in fact complicates the theory. 
In general, if a theory has a number of physical scales, lowering the cutoff and constructing a new EFT at each threshold $\Lambda_i$ of the theory automatically factorizes physics at different distance scales, including the factorization which results from splitting the low-energy degrees of freedom into modes.  SCET is an EFT describing multiple jets of particles in which the invariant mass of pairs of particles within a jet is much less than the invariant mass of any pair of jets. The degrees of freedom of SCET in a given jet are therefore just those of QCD with a UV cutoff $\Lambda$. SCET is more complicated than many canonical EFTs such as four-fermi theory or HQET because of the interactions between the various sectors and the necessity to avoid double counting of degrees of freedom which could be consistently be assigned to more than one sector.

In the formalism presented in \cite{Goerke:2017ioi}, each low invariant mass sector of the theory is described by a different copy of QCD, with interactions between sectors occurring via Wilson lines in the external current. This simplifies the EFT by reducing the number of degrees of freedom and interactions, while also making manifest the scales at which different factorizations occur. It also simplifies the structure of power corrections in the theory, since individual modes in SCET do not manifestly factorize at subleading order due to soft-collinear mixing terms in the Lagrangian\footnote{This factorization was demonstrated at subleading power in \cite{Moult:2019mog}.}, and these are not present in this approach. In addition, since at the matching scale $Q$ the degrees of freedom below $Q$ are not factorized into separate modes, there is no distinction between the EFT for \scetone\ and \scettwo\ processes immediately below $Q$; this distinction occurs at a lower scale where a process-dependent matching onto a soft theory is performed.

In \cite{Goerke:2017ioi}, this approach was demonstrated for a simple \scetone\ observable, DIS in the $x\to 1$ limit, up to subleading order in $1/Q$. The EFT, including operators up to $O(1/Q^2)$,  was renormalized in this framework in \cite{Goerke:2017lei}. It was observed in \cite{Goerke:2017ioi} that it is necessary to subtract the double-counting of low-energy degrees of freedom which are below the cutoff in different sectors. This is required to reproduce the correct cross section at tree level, and is analogous to zero-bin subtraction in SCET \cite{Manohar:2006nz}.  Without this overlap subtraction, ultraviolet divergences in the EFT would be sensitive to the infrared scales of the theory, so the EFT could not be consistently renormalized.

In this paper we consider \scettwo\ observables in the same framework. Soft-collinear factorization in \scettwo\ is quite different from ultrasoft-collinear factorization in \scetone; since the invariant mass of ultrasoft degrees of freedom is parametrically smaller than that of collinear degrees of freedom, ultrasoft-collinear factorization automatically occurs in \scetone\ as the renormalization scale of the EFT is lowered. For example, in DIS the SCET Lagrangian is run from $Q$ down to an intermediate scale $Q\sqrt{1-x}$, at which point the Operator Product Expansion (OPE) of the external current and its conjugate is matched onto a parton distribution function (PDF), effectively integrating the collinear degrees of freedom out of the theory. The matching conditions onto the PDF are the usual jet functions of SCET.

In contrast, soft-collinear factorization is not achieved by lowering the cutoff of the theory, because soft and collinear degrees of freedom have the same invariant mass. In the standard SCET formalism with distinct collinear and soft modes, soft-collinear factorization is required to sum rapidity logarithms of the form $\alpha_s\ln \frac{\mu}{Q} \ln\frac{\mu}{M}$ which arise in \scettwo\ processes, where $Q$, $M$ and $\mu$ are the hard, soft and renormalization scales, respectively. Without resummation, these show up as large logarithms in the matching condition at the scale $\mu\sim M$.  Individual soft and collinear graphs contain rapidity divergences which are unregulated in dimensional regularization.  In order to define the individual graphs, an additional regulator (examples include the $\delta$ regulator \cite{Chiu:2009yx}, the analytic regulator \cite{Smirnov:1997gx,Beneke:2003htl}, the $\eta$-regulator \cite{Chiu:2011qc,Chiu:2012ir}, or the pure rapidity regulator \cite{Ebert:2018gsn}) must be introduced, which allows soft and collinear terms to be factorized in a scheme dependent manner. The scheme dependence introduced by the choice of regulator allows a set of rapidity renormalization group (RRG)  evolution equations to be derived which sum rapidity logarithms \cite{Chiu:2007yn,Chiu:2012ir,Chiu:2011qc,Chiu:2009yx,Chiu:2008vv,Chiu:2007dg}. 

Since the formalism in \cite{Goerke:2017ioi} does not factorize collinear and soft degrees of freedom in the SCET Lagrangian, it is not immediately clear how soft-collinear factorization arises in this approach. As we will show in this paper, rapidity logarithms arise because at the loop level there is an ambiguity in defining the sum of individually divergent contributions from the different sectors of the theory, and the scheme dependence of this ambiguity is analogous to the rapidity cutoff usually introduced to factorize soft and collinear modes in SCET. The ambiguity and corresponding resummation occurs in the matching conditions onto the soft theory, so does not affect the running in the intermediate EFT.

In the next section, we illustrate this with the simplest \scettwo\ process, the massive Sudakov form factor. In the subsequent section we consider the Drell-Yan (DY) process at $q_T^2\ll q^2$. Our conclusions are presented in Sec. \ref{sec:Conclusion}.  

%
\section{The Massive Sudakov Form Factor} \label{sec:formfactor}
%
%

\begin{figure}[h!] \begin{centering}
	\includegraphics[width=0.25\textwidth]{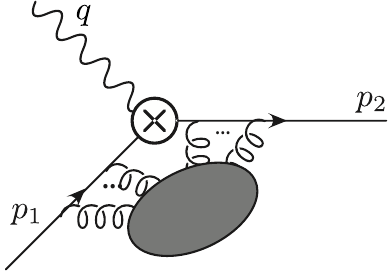}
	\caption{The massive Sudakov form factor.}
\end{centering} \end{figure}
The massive Sudakov form factor  provides a simple example of a physical quantity with rapidity logarithms \cite{Mueller:1989}.
In a theory with a vector boson of mass $M$ the vector form factor $F(Q^2/M^2)$ is defined by
\begin{equation} \label{eq:form_factor_defn}
\bra{p_2} j^\mu \ket{p_1}= F\left(\frac{Q^2}{M^2}\right) \bar{u}_2 \gamma^\mu u_1 \ 
\end{equation}
where 
\begin{equation} \begin{aligned} \label{eq:current_defn}
j^\mu(x)= \bar{\psi}(x) \gamma^\mu \psi(x),
\end{aligned} \end{equation}
and $q^\mu=p_2^\mu-p_1^\mu$ and $Q^2 \equiv -q^2 = 2p_1 \cdot p_2$. The one-loop QCD calculation gives
\begin{equation}
F\!\left(\frac{Q^2}{M^2}\right)=1+\frac{\alphabar}{2}\left( -L_{Q/M}^2  + 3L_{Q/M}  -\frac{4\pi^2}{6} - \frac{7}{2}\right)
\end{equation}
where  $\alphabar\equiv\alpha_s C_F/(2\pi)$ and $L_{Q/M}\equiv\log(Q^2/M^2)$.  The large logarithms of $Q^2/M^2$ in the fixed-order expansion indicate that for $Q^2\gg M^2$, perturbation theory is not well behaved and must be resummed. This is achieved by splitting $F(Q^2/M^2)$ into separate factors, each of which depends only on a single dynamical scale as well as an arbitrary factorization scale; consistency of the factorization formula to all orders in $\alpha_s$ then places sufficient constraints on the perturbative series to allow resummation of the logarithmically enhanced terms to any order in the leading-log expansion.

\subsection{Soft-Collinear Factorization}\label{sec:SCfact}

First we review the standard SCET approach to factorization for this quantity.
In this standard framework, the EFT below $\mu=Q$ is \scettwo\, \cite{Chiu:2009yx} with contributions from $n$-collinear, $\bar n$-collinear and soft (or mass) modes, $p_n\sim Q(\lambda^2,1,\lambda)$, $p_{\bar n}\sim Q(1,\lambda^2,\lambda)$ and $p_s\sim Q(\lambda,\lambda,\lambda)$, where $\lambda\sim M/Q$.
Matching from QCD onto SCET factors out at the hard matching coefficient at the scale $\mu=Q$, giving 
\begin{equation} \begin{aligned} \label{eq:form_factorization}
F\!\left(\frac{Q^2}{M^2}\right)\!&=\left[ 1+\frac{\alphabar}{2}\! \left( -L_Q^2 +3 L_Q  + \frac{\pi^2}{6} - 8 \right) \right] \\ 
&\times  \left[ 1+ \frac{\alphabar}{2}\! \left( -L_M^2 -L_M(3-2L_Q) -\frac{5\pi^2}{6} \!+ \frac{9}{2} \right) \right]\\
\end{aligned} \end{equation} 
where $L_Q=\log(Q^2/\mu^2)$ and $L_M=\log(M^2/\mu^2)$.  The first factor is the hard matching coefficient 
\begin{equation}\label{c2def}
C_2(\mu)= 1+\frac{\alphabar}{2} \left( -L_Q^2 +3 L_Q  + \frac{\pi^2}{6} - 8 \right) 
\end{equation}
from the QCD current to the leading SCET current, and is independent of the infrared scale $M$, while the second factor is the matrix element of the vector current in the effective theory.  

As discussed in \cite{Chiu:2007dg}, the matrix element of the vector current in SCET is problematic, as it has a logarithmic dependence on the ultraviolet scale $Q$, which is above the cutoff of the EFT. Typically in an EFT, logarithms of ultraviolet scales are replaced by logarithms of the cutoff, which allows them to be summed using RGE techniques. As noted in \cite{Becher:2010tm}, the scale $Q$ enters the EFT because the contributions to the loop graph from individual modes are not separately well-defined, so even though  $Q$ is not a dynamical scale associated to any single mode, the sum of the graphs re-introduces $Q$ into the result (this was dubbed the ``collinear anomaly" in \cite{Becher:2010tm}). As a result, integrating the massive gauge boson out of the theory at $\mu=M$ gives matching conditions onto the soft theory containing logarithms of $Q/M$ which are not resummed by the usual RGE evolution.

Rapidity logarithms are resummed in SCET by exploiting an additional scheme dependence in the theory, beyond the choice of renormalization scale $\mu$.  In \scettwo\ processes with rapidity logarithms,  individual collinear and soft graphs are not well-defined; only the sum is. In order to regulate the individual soft and collinear contributions, an additional regulator must be added to the theory. Using, for example, the rapidity regulator of \cite{Chiu:2012ir}, individual soft and collinear contributions are separately well-defined, and the form factor factorizes into individual hard, soft and jet functions,  
\begin{equation}\begin{aligned}\label{scetsudfact}
F\left(\frac{Q^2}{M^2}\right)&=  C_2\left(\mu\right)   S\left(\frac{M}{\nu},\frac{M}{\mu}\right)
\\&\times J_n\left(\frac{p_2^-}{\nu},\frac{M}{\mu}\right)
  J_{\bar n}\left(\frac{p_1^+}{\nu},\frac{M}{\mu}\right)  
\end{aligned}\end{equation}
where $p_1^+$ and $p_2^-$ are the large light-cone components of $p_1^\mu$ and $p_2^\mu$ satisfying $p_1^+ p_2^-=Q^2$, and, to one loop,
\begin{equation}\begin{aligned}\label{scetsudfact2}
J_n\left({p_2^-\over\nu},{M\over\mu}\right)&=1+\alphabar\left[L_M\left(\log{p_2^- \over\nu}-{3\over 4}\right)-{\pi^2\over 6}+{9\over 8}\right] \\
J_{\bar n}\left({p_1^+\over\nu},{M\over\mu}\right)&=1+\alphabar\left[L_M\left(\log{ p_1^+\over\nu}-{3\over 4}\right)-{\pi^2\over 6}+{9\over 8}\right] \\
S\left({M\over\nu},{M\over\mu}\right)&=1+{\alphabar\over2}\left[L_M^2-4 L_M \log{M\over\nu}-{\pi^2\over 6}\right].
\end{aligned}
\end{equation}The rapidity scale $\nu$ defines a scheme-dependent way to separate soft and collinear contributions. While the individual soft and jet functions depend on  $\nu$, their product is $\nu$-independent, thus allowing a renormalization group equation (the rapidity renormalization group) to be derived. Each of the terms may then be evolved from its natural rapidity scale in $\nu$, summing the rapidity logarithms.

Similar results have also been derived in the collinear anomaly formalism \cite{Becher:2010tm,Becher:2011jb,Becher:2012njb}, in which the product of $J_n J_\nbar S$ in \eqn{scetsudfact} is re-factorized as the product of two functions: an anomaly exponent $F$ in which the rapidity logarithms appear and a remainder function $W$ independent of the hard scale. 

%
\subsection{Scheme Dependence Without Modes} \label{sec:ambiguity}
%
%
In the formalism introduced in \cite{Goerke:2017ioi}, there are no explicit modes, so rapidity logarithms are not resummed by exploiting the scheme dependent separation into soft and collinear degrees of freedom.  Instead, as we now discuss, the contributions from the individual $n$ and $\nb$ sectors of the theory, along with the corresponding overlap subtraction, are individually divergent, and the scheme dependence in defining their sum allows rapidity logarithms to be summed.

In this formalism, the incoming and outgoing states are each described by two decoupled copies of QCD. Each sector interacts with the other sector as a lightlike Wilson line, contained in the hard external current, since gluons with sufficient momentum to deflect the worldline of the other sector have been integrated out of the theory.  While the theory is frame-independent, for simplicity we work in the Breit frame and label the sectors by the light-like directions $n^\mu=(1,0,0,1)$ and $\bar n^\mu=(1,0,0,-1)$, with the light-cone coordinates of a four-vector $p^\mu$ defined as $p^+\equiv p\cdot n$, $p^-\equiv p\cdot\bar n$.  The incoming quark is in the $\nb$-sector, $p_1^+\gg (p_1^-, |p_{1\perp}|)$, while the outgoing is in the $n$-sector, $p_2^-\gg (p_2^+, |p_{2\perp}|)$.

At leading order, the hard QCD current matches onto the scattering operator $O_2$ via the matching relation
\begin{equation} \begin{aligned} \label{eq:current_expansion}
j^\mu(x) \to j^\mu_{\rm SCET}= C_2(\mu) O_2^\mu(x) + O\left( \frac{1}{Q} \right) \ ,
\end{aligned} \end{equation} 
where $C_2(\mu)$ is given in \eqn{c2def}, and the neglected subleading operators are known up to order $1/Q^2$ when there are two sectors \cite{Goerke:2017lei}.
The operator $O_2^\mu(x)$ is defined as 
\begin{equation} \begin{aligned} \label{eq:O2defn}
O_2^\mu(x)&= [\bar{\psi}_n(x_n) \QuarkWinc_n(x_n)] P_\nb \gamma^\mu P_\nb [\QuarkWout_\nb(x_\nb) \psi_\nb(x_\nb)]  
\end{aligned} \end{equation} 
where the fields $\psi_n$ and $\psi_{\bar n}$ are QCD quark fields in the two sectors, and 
\begin {equation}
P_n={\ns \nbs\over4},\ P_{\nb}={\nbs\ns\over 4}.
\end{equation}
The square brackets separate the field content of each sector.  The (un)barred Wilson lines are (outgoing) incoming, and are defined \cite{Chay:2004zn,Arnesen:2005nk,Feige:2013zla}) as 
\begin{equation} \begin{aligned} \label{eq:WlineDefn}
\QuarkWinc_n(x) &= P \exp \left(\,ig\int_{-\infty}^0 ds\, \nb \cdot A_n(x+\nb s) e^{ s0^+} \right) \\
\QuarkWout_\nb(x) &= P \exp \left(\,ig\int_0^{\infty}  ds\, n \cdot A_\nb(x+n s) e^{-s0^+}\right)
\end{aligned} \end{equation} 
where again the subscript in the gluon fields $A^\mu_{n,\bar n}$ labels the sector. Note that we are using the labelling convention that $W_\nbar^\dagger$ is a Wilson line along the $n$ direction, coupling to fields in the $\nbar$ sector.

Finally, consistently expanding the QCD amplitude in powers of $1/Q$ also means that the energy-momentum conserving delta function must also be expanded, giving
\begin{equation}\begin{aligned}\label{eq:multipole}
\delta_{\rm SCET}(Q;p_n, p_{\bar n})\equiv &\ 2\delta(p_n^--Q^-)\delta(p_{\bar n}^+-Q^+) 
\\&\times \delta(\myvec p_{nT}+\myvec p_{\bar nT}-\myvec q_T) + \dots
\end{aligned}\end{equation}
where $p_n$ and $p_{\bar n}$ are the total momenta in the $n$ and $\nb$ sectors, respectively. 
This is achieved by multipole expanding the $x^\mu$ dependence of the current in \eqn{eq:O2defn}, where we have defined
\begin{equation}
x_n^\mu \equiv x^+ \frac{\nb^\mu}{2} + x_\perp^\mu,\ \ x_\nb^\mu \equiv x^- \frac{n^\mu}{2} + x_\perp^\mu.
\end{equation}
Multipole expanding the energy-momentum conserving delta function has no effect on the renormalization of $O_2$ since the sectors are decoupled, but ensures correct power counting when calculating production rates, as we will see in the next section for Drell-Yan production.

\begin{figure}[t!] 
   \centering
  \includegraphics[width=0.45\textwidth]{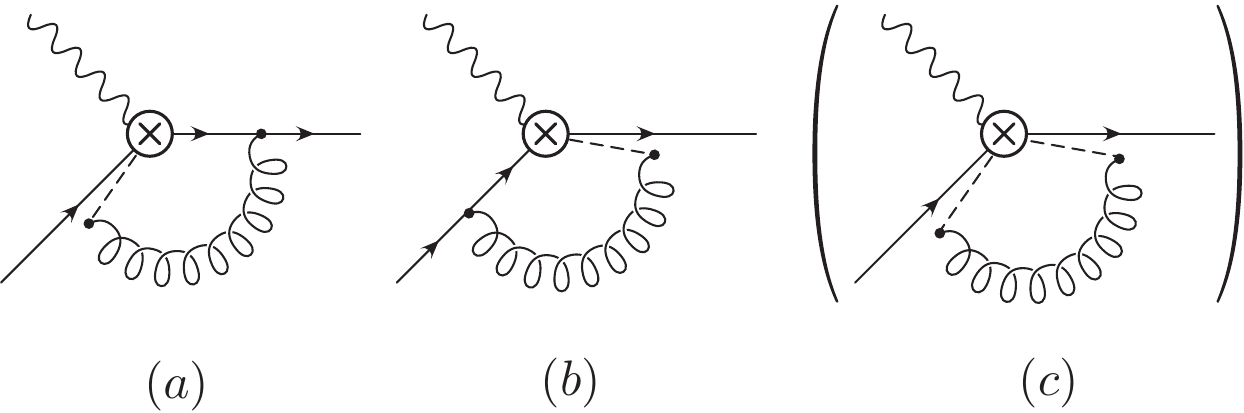}
   \caption{Renormalization of $O_2$.  Diagram (c) is the overlap subtraction. In (a), the gluon is in the $n$ sector; in (b), it is in the $\nb$ sector.  In (c), the dashed lines represent Wilson lines $W_{n,\nb}$, depending on their direction. }
   \label{o2oneloop}
\end{figure}

As described in \cite{Goerke:2017ioi}, this theory double counts quarks and gluons whose momentum is below the cutoff of both sectors, and the effects of this double counting must be explicitly subtracted from diagrams. This ``overlap subtraction" is similar to the familiar zero-bin subtraction in SCET \cite{Manohar:2006nz}, or the equivalent soft subtraction prescription discussed in \cite{Lee:2006nr,Idilbi:2007ff,Idilbi:2007yi}. At tree level it is required to ensure that external states are not double counted in the rate. At one loop this corresponds to subtracting the overlap graph in Fig. \ref{o2oneloop}(c), which is equivalent to either the $n$- or $\nbar$-sector graph, but with the quark propagator replaced by the corresponding lightlike Wilson line. Formally this corresponds to dividing matrix elements of $O_2$ by the vacuum expectation value of Wilson lines,
\begin{equation} \begin{aligned} \label{eq:amplitude_subtraction}
\bra{p_2} O_2(x) \ket{p_1}_\mathrm{subtracted}= \frac{ \bra{p_2} O_2(x) \ket{p_1} } { \frac{1}{N_C} \mathrm{Tr} \bra{0} W_\nbar^\dagger(x) \overline{W}_n(x)\ket{0}  }  .
\end{aligned} \end{equation} 
This prescription means that the one-loop matrix element of $O_2$ is given by the combination 
\begin{equation} \begin{aligned}
\Mmt_1=\Gamma_n+\Gamma_{\bar{n}}-\Gamma_{\sub}-2  \frac{\Gamma_\psi}{2} \ ,
\end{aligned} \end{equation} 
where the $\Gamma_i$ represent the one-loop $n$-sector, $\bar{n}$-sector, and overlap subtraction graphs in Fig.
 \ref{o2oneloop}(a), (b) and (c), and $\Gamma_\psi$ is  the wavefunction renormalization contribution,
\begin{equation} \begin{aligned}
\Gamma_\psi=\frac{1}{2}\alphabar \Mmt_0 \bigg(\frac{1}{\epsilon}  - L_M- \frac{1}{2}\bigg) 
\end{aligned} \end{equation} 
where  $\Mmt_0=\bar{u}_2 P_{\bar{n}} \gamma^\mu P_{\bar{n}} v_1$ and we work in $d=4-2\epsilon$ dimensions.

As described in \cite{Goerke:2017ioi} (and, in a different context, \cite{Chiu:2009yx,Idilbi:2007ff,Freedman:2014uta}), while the terms $\Gamma_n$, $\Gamma_{\bar n}$ and $\Gamma_{\sub}$ are all individually divergent even when the theory is regulated in dimensional regularization, adding together the individual graphs before doing the final momentum integral results in a finite answer in $d$ dimensions:
\begin{equation} \begin{aligned}  \label{eq:naiveadd}
\Mmt_1 =&\frac{\alphabar}{2}\Mmt_0 \bigg[ \frac{2}{\epsilon^2}+ \left({1\over\epsilon}-L_M\right)\left(3-2L_Q\right) 
\\& -L_M^2  -{5\pi^2\over 6} + \frac{9}{2}\bigg]. 
\end{aligned} \end{equation} 
After adding the appropriate counterterm, this reproduces the second line in \eqn{eq:form_factorization}. This result was used in \cite{Goerke:2017ioi} to define the one-loop renormalization of $O_2$ in this formalism.

However, one must be careful here, because na\i\" vely adding together divergent graphs is not a well-defined procedure. In particular, adding the integrands before performing the final integration corresponds to only one possible scheme to define the sum of the divergent graphs.
We can illustrate this scheme dependence by doing the $k^+$ integrals for $\Gamma_{n,\nb,\rm{sub}}$ by contours for each graph and then doing the $(d-2)$-dimensional
$k_\perp$ integrals, but leaving the divergent $k^-$ integrals unevaluated.  
This gives for the $n$-sector graph, Fig.\ \ref{o2oneloop} (a),
\begin{equation} 
\Gamma_n =  
C_\epsilon\int_{0}^{p_2^-} \! \frac{dk^-}{-k^-} \left( 1- \frac{k^-}{p_2^-} \right)^{1-\epsilon} \end{equation}
where 
\begin{equation}
C_\epsilon\equiv\alphabar \Mmt_0 \left( \frac{\mu^2 e^{\gamma_E}}{M^2}\right)^\epsilon \Gamma(\epsilon).
\end{equation}
The $\nb$-sector graph gives
\begin{equation} \begin{aligned} 
\Gamma_{\bar{n}}&= C_\epsilon\left[ \frac{1}{1-\epsilon}+
\int_{0}^{\infty} \! \frac{p_1^+ dk^-}{M^2} \frac{ 1 - \left(\frac{p_1^+ k^-}{M^2}\right)^{-\epsilon} }{1-\frac{p_1^+ k^-}{M^2}} \right] \\
&= C_\epsilon 
\Bigg[
    \frac{1}{1-\epsilon}+ \pi \csc(\pi\epsilon) (-1+i0^+)^{-\epsilon} 
    \\&\qquad+
    \int_{0}^{\infty} \! \frac{p_1^+ dk^-}{M^2} \frac{ 1 }{1 - \frac{p_1^+ k^-}{M^2}}
\Bigg]  
\end{aligned} \end{equation} 
and finally the overlap subtraction graph gives
\begin{equation} 
\Gamma_{\sub}= C_\epsilon \int_{0}^{\infty} \! \frac{dk^-}{-k^-}.
\end{equation} The $n$-sector and $\nb$-sector graphs are divergent as $k^-\to 0$ and $k^-\to\infty$, respectively. When the graphs are added together before doing the final integral, these divergences are cancelled by the overlap graph, giving the result in \eqn{eq:naiveadd}. However, the individual terms $\Gamma_i$ each arise from loops containing distinct particles in the EFT ($n$- and $\nb$-collinear gluons and their overlap), so we are free to individually rescale the momenta in the individual integrals before combining them. For example, rescaling the integration variable in $\Gamma_\nb$ by
$k^-\to {\zeta^2\over Q^2} k^-$
will instead give the sum of the three graphs
\begin{equation}\begin{aligned}
\int_0^{p_2^-}\!\! &\left[{1\over -k^-}\left( 1- \frac{k^-}{p_2^-} \right)^{1-\epsilon}
\!\!\!\!+{\zeta^2 p_1^+\over M^2 Q^2}{1\over 1-{k^- \zeta^2 p_1^+\over M^2 Q^2}}+{1\over k^-} \right]dk^-\\
& +\int_{p_2^-}^\infty \left[{\zeta^2 p_1^+\over M^2 Q^2}{1\over 1-{k^- \zeta^2 p_1^+\over M^2 Q^2}}+{1\over k^-} \right]dk^-\\
&= 1+\log{M^2\over\zeta^2}+i\pi+\left(1-{\pi^2\over 6}\right)\epsilon +O(\epsilon^2) \equiv I(\zeta)
\end{aligned}\end{equation}
which gives the $\zeta$-dependent matrix element 
\begin{equation} \begin{aligned}  \label{eq:mat_elem_zeta}
\Mmt_1  =&\frac{\alphabar}{2}\Mmt_0 \bigg[ \frac{2}{\epsilon^2}+ \left({1\over\epsilon}-L_M\right)\left(3-2L_\zeta\right) \\& -L_M^2  -{5\pi^2\over 6} + \frac{9}{2}\bigg] 
\end{aligned} \end{equation} 
where $L_\zeta=\log( \zeta^2/\mu^2)$. Choosing $\zeta=Q$ corresponds to the na\"\i ve result (\eqn{eq:naiveadd}), but leaving $\zeta$ free makes the scheme dependence manifest. This also underscores the fact that SCET has no dynamical dependence on the scale $Q$, which has been integrated out of the theory: the $Q$ dependence in the na\"\i ve matrix element is in fact $\zeta$ dependence, which parameterizes the scheme-dependence of the rapidity divergent integrals. A similar calculation was performed with massless gluons in \cite{Idilbi:2007ff}
where the authors noted that the scaleless SCET integrals had the scale $Q$ inserted by hand; any other scale $\zeta$ could similarly be inserted by hand, but the choice $\zeta=Q$ was ``justified \emph{a postieri} by the requirement that SCET reproduce the IR divergences of QCD''. 

This simple one-loop example demonstrates how rapidity logarithms of the hard scale $Q$ enter into the EFT: they are not logarithms of $Q$ in matrix elements, but rather logarithms of some dimensionful scheme parameter which defines how individually rapidity divergent graphs in different sectors are added together. The scheme dependence of the matrix element in \eqn{eq:mat_elem_zeta} suggests that we introduce a corresponding scheme dependence in the one-loop matching coefficient from QCD onto SCET,
\begin{equation} \begin{aligned}\label{O2matchamb}
C_2(\mu&)\to C_2( \mu,\zeta)\\ &
=1+\frac{\alphabar}{2} \left( -L_Q^2 +3 L_Q +2L_M L_{Q/\zeta} +{\pi^2\over 6} - 8\right)
\end{aligned} \end{equation} 
where $L_{Q/\zeta}\equiv\log{Q^2/\zeta^2}$, so that physical quantities are independent of $\zeta$. However, as we will discuss in detail in the next section, the Wilson coefficient $C_2$ must be independent of the infrared scale $M$, which requires choosing $\zeta=Q$ at the matching scale $\mu=Q$, eliminating the nonanalytic dependence on $M$ in \eqn{O2matchamb}.  Thus, it would seem that there is no freedom to choose $\zeta$ in SCET, since it is fixed to $\zeta=Q$ by the requirement that the scales $M$ and $Q$ factorize. However, the fact that logarithms of $Q$ in matrix elements of $O_2$ are in fact logarithms of a scheme parameter allows us to sum rapidity logarithms in low-energy matrix elements. Since the $\zeta$-scheme defined in this section was introduced for illustrative purposes and is not obviously defined beyond the simple one-loop graphs considered here, we will discuss resummation of rapidity logarithms with a well-defined regulator in the next section.

%
\subsection{Resummation} \label{sec:resummation}
%

There are a number of regulators in the literature which regulate rapidity divergences
\cite{Collins:2011zzd,Chiu:2009yx,Smirnov:1997gx,Li:2016axz,Chiu:2012ir,Ebert:2018gsn}; the most instructive for our purposes is to use a version of the $\delta$-regulator \cite{Chiu:2009yx}. In its original formulation, 
quark propagators and Wilson lines were both modified by adding a quark mass term to the Lagrangian and using the new quark propagator to derive the new Wilson line propagator. Here we leave quark and gluon propagators unchanged and simply redefine the Wilson lines by shifting the pole prescription $i0^+ \rightarrow -\delta_n+i0^+$ for both the $\overline W$ and $W^\dagger$ in \eqn{eq:WlineDefn},
and we allow each sector label $n_i$ to have a separate value of $\delta$ (i.e. $\delta_\nb$ in the $\nb$-sector and $\delta_o$ in the overlap between the sectors\footnote{This differs from the prescription in \cite{Goerke:2017ioi}, where the overlap Wilson lines had the same value of $\delta$ as the corresponding sector.}).
With this modification, the $n$-sector graph becomes

\begin{equation}
\Gamma_{n}^{\delta} =  \alphabar \Mmt_0 \left[ \left(\frac{1}{ \epsilon} - L_M \right) \left(\log\frac{\delta_n}{p_2^-} + 1 \right) - {\pi^2\over 6}  + 1 \right] \ ,
\end{equation} 
and the $\nb$ sector gives the same result but with $\delta_n \rightarrow \delta_\nb$ and $p_2^- \rightarrow p_1^+$. The overlap graph contributes
\begin{equation}
\Gamma_{\sub}^{\delta} =  \alphabar \Mmt_0 \left[ -\frac{1}{\epsilon^2 } 
+ \left( \frac{1}{ \epsilon} - L_M\right) \log\frac{\delta_o^2}{\mu^2} + \frac{L_M^2 + {\pi^2\over 6}}{2}  \right] \ ,
\end{equation} 
so that together with the wavefunction graphs we find 
\begin{equation} \begin{aligned}\label{M1delta}
\mathcal{M}_{1}^{\delta} =  &\frac{\alphabar}{2} \Mmt_0 \bigg[ \frac{2}{\epsilon^2 } + \left( \frac{1}{ \epsilon} - L_M\right) 
\left(3-2\log {\nu^2\over\mu^2} \right) 
\\& - L_M^2 +\frac{9}{2} - {5\pi^2\over 6}    \bigg] \ ,
\end{aligned}\end{equation}
where the parameter $\nu$, defined by
\begin{equation}
{\delta_n\delta_\nb\over\delta_o^2}\equiv{ Q^2\over\nu^2} ,\end{equation}
plays a role analogous to $\zeta$ in the previous section.
We can take the regulators $\delta_i$ to zero while keeping $\nu$ fixed, and the scheme-dependence of the rapidity log is then reflected in the $\nu$ dependence of the result.\footnote{The scheme dependence in taking the $\delta_i\to 0$ limit was also stressed in \cite{JCollins:2012tmdpdf}.}  We note that $Q$ as introduced here is not a dynamical scale in the EFT, but simply serves to define the dimensionful parameter $\nu$.

Setting $\nu=Q$ in \eqn{M1delta} gives the na\"\i ve result, \eqn{eq:naiveadd}.
More generally, the scheme dependence of the matrix element in \eqn{M1delta} requires a corresponding scheme dependence in the Wilson coefficient of $O_2$ so that physical quantities are independent of $\nu$. We must be careful, however, in defining the rapidity scheme. It is a general feature of EFTs that Wilson coefficients do not have nonanalytic dependence on infrared scales; otherwise, the EFT would not factorize the physics of short and long distance scales. At scales $\mu$ parametrically larger than $M$, this can only be achieved by choosing $\nu=Q$; otherwise the Wilson coefficient $C_2(\mu)$ would contain a factor of $L_M\log{Q^2/\nu^2}$, which is sensitive to the IR scale $M$. As with the $\zeta$-scheme in the previous section, it would therefore seem that there is no freedom to choose the rapidity regulator in SCET, since it is fixed to $\nu=Q$ by the requirement that the scales $M$ and $Q$ factorize. However, after running the theory down to the scale $\mu=\mu_S\sim M$, the gluon mass is no longer an infrared scale, and we are then free to run $\nu$ from $\nu_H\equiv Q$ to $\nu_S=M$ when calculating the matching conditions onto the free theory, summing the rapidity logarithms in the matching condition at $\mu_S$.

After running the matching coefficient $C_2$ from $\mu=Q$ down to $\mu_S\sim M$, we integrate out the massive gluon and match $O_2$ onto a free theory,
\begin{equation}\label{softmatch1}
O_2^\mu(x)\to C_S O_S^\mu(x)
\end{equation}
where 
\begin{equation}
O_S^\mu(x) = \bar \psi(x)P_\nb \gamma^\mu P_\nb \psi(x)
\end{equation}
and the $\psi$'s are free fermions. However, the resulting matching coefficient
\begin{equation} \begin{aligned}\label{softmatch2}
C_S = 1+  \frac{\alphabar}{2} \bigg[  -L_M\left(3-2\log{\nu_H^2\over\mu_S^2}\right) 
-L_M^2 -{5\pi^2\over 6} + \frac{9}{2}\bigg]
\end{aligned} \end{equation}
contains a large rapidity logarithm. To resum this, we must effectively run the matching condition $C_S$ in rapidity from $\nu$ from $\nu=\nu_H$ to $\nu=\nu_S\sim \mu_S$ before integrating out the gluon, which we do by running the operator $O_2$ in rapidity at the matching scale $\mu_S$. We define
\begin{equation}\label{o2match}
\left. C_2 O_2^\mu(x)\right|_{\mu=\mu_S}= C_2\left(\mu_S\right) V_J\left({\mu_S\over M},{\nu\over Q}\right) O_2^\mu(x,\nu)
\end{equation}
where $O_2^\mu(x,\nu)$ denotes $O_2(\mu_S)$ defined with $\nu\neq Q$, and
at one loop,
\begin{equation} \begin{aligned} \label{CJleading}
V_J\left({\mu\over M},{\nu\over Q}\right)&
=1+ \alphabar L_M\log\frac{Q^2}{\nu^2}. 
\end{aligned} \end{equation} 
The fact that the $Q$ dependence of $O_2$ factorizes according to \eqn{o2match} means that the logarithm of $Q$ in \eqn{CJleading} exponentiates. Explicitly, differentiating \eqn{o2match} with respect to $\log\nu$ gives the equation
 \begin{equation} \begin{aligned}  \label{eq:evolution_cf_cs}
\frac{d}{d\log\nu} V_J&=\left(-\langle O_2^\mu(x,\nu)\rangle^{-1}\frac{d}{d\log\nu}\langle O_2^\mu (x,\nu)\rangle\right) V_J\\ &\equiv \gamma_\nu^J V_J \\
\end{aligned} \end{equation} 
where 
\begin{equation} \begin{aligned}  \label{eq:anomdim_cf_cs}
\gamma_\nu^J &= -2\alphabar L_M. 
\end{aligned} \end{equation} 
This has the solution
\begin{equation} \begin{aligned}  \label{eq:unitary_Vf}
\log V_J\left(\frac{\mu_S}{M},{\nu_S\over Q}\right)  = \int_{Q}^{\nu_S} \! \frac{d\nu}{\nu} \gamma_\nu^J = \alphabar(\mu_S) \log\frac{M^2}{\mu_S^2} \log\frac{Q^2}{\nu_S^2}
\end{aligned} \end{equation} 
which corresponds to running the rapidity scale from $\nu=Q$ to $\nu_S$.

Having resummed the large rapidity logarithms into $V_J$, the heavy gauge boson is then integrated out, and Eqs.\ (\ref{softmatch1}) and (\ref{softmatch2})  become
\begin{equation}
O_2^\mu\left(x,\nu_S\right)\to C_S\left({\mu_S\over M},{\nu_S\over \mu_S}\right) O_S(x)
\end{equation}
and
\begin{equation} \begin{aligned}
C_S\left({\mu\over M},{\nu\over\mu}\right) = 1+  &\frac{\alphabar}{2} \bigg[  -L_M\left(3-2\log{\nu^2\over\mu^2}\right) 
\\&-L_M^2 -{5\pi^2\over 6} + \frac{9}{2}\bigg]
\end{aligned} \end{equation}
which has no large logarithms at $\mu\sim\nu\sim M$.
We can then combine all of these steps to obtain the resummed factorization formula 
%
\begin{equation} \begin{aligned} \label{eq:sudakov_final_factorization}
F\left( {Q^2\over M^2} \right)
 &= U_2(\mu_S,\mu_H)C_2\left( \mu_H \right)   
 \\&\times
 V_J\left({\mu_S\over M},{\nu_S\over Q} \right) C_S\left({\mu_S\over M},{\nu_S\over \mu_S} \right)\end{aligned} \end{equation}
where 
\begin{equation} \begin{aligned} \label{eq:U2evolution}
\log U_2( \mu_S , \mu_H) &= \int_{\mu_H}^{\mu_S} {d\mu \over \mu} \gamma_2^{LL}(\mu) 
\\&=\frac{4 C_F^2}{\beta_0^2}\bigg[ \frac{1}{\alphabar(\mu_H)} -\frac{1}{\alphabar(\mu_S)} 
\\& \qquad\qquad -\frac{1}{\alphabar(Q)}\log{\left(\frac{\alphabar(\mu_S)}{\alphabar(\mu_H)}\right)}\bigg]
\end{aligned} \end{equation}
is the usual leading-log renormalization group evolution of $C_2$  \cite{Bauer:2003pi,Bosch:2004th,Neubert:2004dd}, and
\begin{equation} \begin{aligned} \label{eq:U2evolution_2}
\gamma_2^{LL}(\mu) = 2 \bar\alpha(\mu)\log\frac{Q^2}{\mu^2} \equiv \Gamma_\mathrm{cusp}[\bar\alpha]\log\frac{Q^2}{\mu^2} \ .
\end{aligned} \end{equation}
This reproduces the results of  \cite{Chiu:2012ir}, with the caveat that, since this formalism explicitly performs the RRG at the scale $\mu=\mu_S\sim M$, logarithms of $\mu/M$ which are resummed in the expression $\log{\alpha_s(\mu)\over\alpha_s(M)}$ in $\log V_J$ in \cite{Chiu:2011qc} do not require resummation here.

While \eqn{eq:sudakov_final_factorization} is equivalent to the factorization formula in \eqn{scetsudfact}, it arises differently in this form of the EFT. In \eqn{scetsudfact}, the $J_i$ are matrix elements of collinear fields; here $V_J$ corresponds to the rapidity evolution factor of $O_2(x,\nu)$. The assignment of factors of $L_M$ and constants to $O_J$ and $O_S$ also differs from that of \eqn{scetsudfact2}, and more closely resembles the refactorized form of \cite{Becher:2011jb,Becher:2012njb}, but the particular arrangement of these terms is irrelevant for summing logarithms since $\alpha_s$ is evaluated at the same scale in both the soft and jet functions in \scettwo\ processes. We could also choose to define separate rapidity scales for each  sector, $\nu_n\equiv p_1^+ \delta_o/\delta_n$ and $\nu_\nb\equiv p_2^-\delta_o/\delta_\nb$, which would then allow  us to write $V_J$ as the product of two separate factors, in direct analogy with the two jet functions of \eqn{scetsudfact2}; however, this is not necessary for the present case, where the rapidity scales always appear as the product $\nu_n \nu_\nb =\nu^2$.

There are also some important differences between the rapidity running of $O_2(x,\nu)$ and the rapidity renormalization group of \cite{Chiu:2012ir}.  In \cite{Chiu:2012ir},  separate soft and collinear contributions to $O_2$ are defined and separately run in rapidity space; the regularization scheme is defined so that the product of soft and collinear factors is regulator-independent. In our case, matrix elements of operators in the EFT have explicit dependence on the rapidity regulator, which is cancelled by the regulator dependence of the corresponding Wilson coefficient $C_2 V_J$ in the EFT. In addition, since the rapidity regulator introduces sensitivity to the matching scale $\mu_S\ll Q$ into the Wilson coefficient of $O_2$, the variation of $\nu$ is performed at the matching scale $\mu_S$, not at a higher scale. There is a physical reason for this: unlike the renormalization group running of $O_2$ in $\mu$, rapidity running is not universal in SCET, but depends on the particular process of interest. In Drell-Yan at low $q_T^2$, for example, and as discussed in the next section, the rapidity logarithms arise in the matching conditions of products of $O_2$ onto products of parton distribution functions, and are distinct from the rapidity logarithms in \eqn{o2match}. Thus, in this formalism, in which the same SCET Lagrangian may be used to calculate a variety of observables with different rapidity logarithms (or none at all), rapidity logarithms arise in low-energy matching coefficients and are resummed at the appropriate matching scale.

This is also apparent from \eqn{eq:unitary_Vf}: the resummed rapidity logarithms are all multiplied by factors of $\alpha_s(\mu_S)$, so rapidity evolution naturally occurs at the low matching scale. This is also the case using the usual SCET RRG formalism: although in \cite{Chiu:2012ir} it was shown that one can evolve along any path in the $(\mu,\nu)$ plane to obtain the resummed factorization formula, performing rapidity running away from $\mu=\mu_S$ requires an additional resummation of the large logarithms $L_M$ in the rapidity anomalous dimensions of the jet and soft functions in order to achieve an equivalent result.

Just as in the usual RRG formalism, consistency of the factorization (\ref{eq:sudakov_final_factorization}) places constraints on the rapidity anomalous dimension $\gamma_\nu^J$. In \cite{Chiu:2012ir} these constraints were derived using independence of path in the $(\mu,\nu)$ plane; we obtain analogous results by requiring consistency between evolving to two different soft scales $\mu_S$ which differ by order 1. The difference in $V_J$ evaluated at two different soft scales $\mu_S$ and $\mu_S^\prime$ is of order $\log{Q^2/M^2}$, and so contains a large logarithm of $Q$. Since the overall variation of the form factor with respect to $\mu_S$ must vanish, and since matrix elements of $O_2$ are independent of $Q$, this large variation of $V_J$ must be compensated for in the running of $C_2(\mu)$. This means that the change in $C_2 V$ resulting from varying $\mu_S$ by an amount of order 1 is $Q$ independent, which implies
\begin{equation}
    {d\over d\log Q}{d\over d\log\mu}\left(\log U_2+\log V_J\right)=0.
\end{equation}
Since ${d\log U_2\over d\log\mu}=\gamma_2^{LL}$ and ${d\log V_J\over d\log Q}=-{d\log V_J\over d\log\nu}=-\gamma_\nu^J$, this immediately gives the relation
\begin{equation}
    {d \gamma_2^{LL}\over d\log Q}={d\gamma_\nu^J\over d\log\mu}=2\Gamma_{\rm cusp}.
\end{equation}
%
%
\section{Drell-Yan at Small \texorpdfstring{$\mathbf{q_T}$}{qT}}\label{sec:DY}
%
%

A somewhat more involved process with rapidity logarithms is Drell-Yan (DY) scattering, $N_1(p) N_2( \bar p ) \rightarrow \gamma^* +X \rightarrow (\ell \bar{\ell}) + X$, with $q_T^2\ll q^2$, where $q^\mu$ and $q_T^\mu$ are the total and transverse momenta of the final state leptons, respectively. In standard SCET, this is a \scettwo\ process in which the product of two hard external currents may be written in terms of a convolution of transverse-momentum dependent parton distribution functions (TMDPDFs) or beam functions 
\cite{Stewart:2009yx,Becher:2010tm, Echevarria:2012qe, Chiu:2012ir,JCollins:2012tmdpdf,Chay:2012mh} with $n$-collinear, $\nb$-collinear and soft modes, which individually exhibit rapidity divergences. 
By running the TMDPDFs using both the usual $\mu$-renormalization group and its counterpart in rapidity space, logarithms associated with ultraviolet divergences and rapidity divergences may both be summed. If $q_T\equiv \sqrt{\myvec{q}_T^{\,2}}\gg\lqcd$, an additional expansion may then be performed in powers of $\lqcd/q_T$, allowing the product of TMDPDFs to be matched onto the usual parton distribution functions (PDFs). 

Proceeding in an analogous fashion to the previous section, in our formalism the QCD current is first matched at the hard matching scale onto the corresponding SCET current in a theory with $n$ and $\nb$ sectors with the appropriate overlap subtractions. Again,  there is no distinction between \scetone\ and \scettwo\ at the hard matching scale, since amplitudes are expanded in powers of $p^+$, $\bar p^-$, $p_\perp$ and $\bar p_\perp$, with no hierarchy assumed between these scales. The EFT is then evolved via the RGE until a scale $\mu=\mu_S\sim q_T$, at which point the product of currents is matched onto a convolution of PDFs in the soft theory below $\mu=\mu_S$. However, the rate given by this product of currents has an integration ambiguity similar to that discussed in the previous section, and so rapidity logarithms arise when evaluating the matching conditions onto the soft theory. These may be summed by first matching the product of the currents at $\mu=\mu_S$ onto a nonlocal operator equivalent to a convolution of TMDPDFs, which is then run in rapidity space before matching onto PDFs.

%
\subsection{Factorization by Successive Matching}
%

The DY process is mediated in SCET by the dijet current, defined by the matching relation
\begin{equation} \begin{aligned} \label{eq:O2primedefn}
j^\mu(x)&=C_2(\mu)O_2^\mu(x)+\dots \\&= C_2(\mu)\left[ \bar{\psi}_\nbar(x_\nbar) \overline W_\nbar(x_\nbar)    \right]  \\&\qquad\times P_n \gamma^\mu P_n \left[ \overline W_n^\dagger(x_n) \psi_n(x_n) \right] +\dots
\end{aligned} \end{equation} 
where the ellipses denote power corrections, and
\begin{equation}
\QuarkWinc_n^\dagger(x) = \overline P \exp \left[-ig\int_{-\infty}^0 ds\, \nb \cdot A_n(x+\nb s) e^{ s0^+} \right].
\end{equation}
We have denoted the dijet operator as $O_2$ as in the previous section, but in this case both the quark and antiquark are incoming, so the two Wilson lines are also incoming. As in the previous section, the coordinates $x_n$ and $x_\nb$ give the expanded energy-momentum conserving delta functions in \eqn{eq:multipole}, which gives the correct power counting for $q_T^2\ll q^2$. 
The differential cross section for DY is then proportional to the sum over states
\begin{equation}\label{sumX}\begin{aligned}
\sum_{X}\bra{N_1(p) N_2(\bar p)} O_2^{\mu\dagger}\ket{X}\bra{X}O_{2\mu}  \ket{N_1(p) N_2(\bar p)}.
\end{aligned} \end{equation}
Following the standard derivation, we perform the sum over states in \eqn{sumX}, then color- and spinor-Fierz the product of currents into the form \cite{Becher:2010tm,Becher:2015gsa,Collins:1984kg,Bauer:2010vu}

\begin{equation}\label{fierz}
O_2^{\mu\dagger}(x)O_{2\mu}(0)=-\frac{1}{N_c}\left[\bar\chi_n(x_n){\nbs\over 2}\chi_n(0)\right]\left[\bar\chi_\nb(0){\ns\over 2}\chi_\nb(x_\nb)\right]
\end{equation}
where $\chi_{n}\equiv \overline W^\dagger_{n}\psi_n$, $\bar{\chi}_n = \bar{\psi}_n \overline{W}_n$, and similar for $\chi_\nb$. We have neglected terms which vanish when taking color- and spin-averaged matrix elements.
The differential rate may then be written
\begin{widetext}
\begin{equation}\begin{aligned}\label{dsigma}
d\sigma&={4\pi\alpha^2\over 3 q^2 s}{d^4 q\over (2\pi)^4} (-g_{\mu\nu}) |C_2(\mu)|^2 \int d^4 x \ e^{-iq\cdot x} \langle N_1(p) N_2(\bar p)|O_2^{\mu\dagger}(x)O_{2}^{\nu}(0)|N_1(p) N_2(\bar p)\rangle\\
&={4\pi\alpha^2\over 3N_cq^2 s}{dq^+ dq^- d^2\myvec q_T}  |C_2(\mu)|^2 \langle N_1(p) N_2(\bar p)|\ttwo(q^-, q^+, \myvec q_T)|N_1(p) N_2(\bar p)\rangle\\
\end{aligned}
\end{equation}
\end{widetext}
where 
$s=(p+\bar p)^2$, and the non-local operator
\begin{equation}\label{t22def}
\begin{aligned}
\ttwo(q^+\!, q^-\!, \myvec q_T)\!=\!&\int\! \frac{d \xi_1 d\xi_2}{(2\pi)^d} d^{d-2}\myvec x_T \, e^{-i\xi_1 q^-}\!e^{-i\xi_2 q^+} e^{i\myvec q_T\cdot \myvec x_T}
\\&\times\left[ \bar{\chi}_n(\nb\xi_1+\myvec x_T)\frac{\slashed{\nb}}{2} \chi_n(0) \right] 
\\&\times \left[   \bar{\chi}_\nb (0)\frac{\slashed{n}}{2} \chi_\nb(n \xi_2+\myvec x_T) \right]
\end{aligned}
\end{equation}
is the Fourier transform of the product of position-space TMDPDFs (defined here in $d$ dimensions).
This has the same form as Equation (9) of \cite{Becher:2010tm}, although in our case the theory in which $\ttwo$ is defined does not have separate collinear and soft modes. Note that at this stage the introduction of $\ttwo$ is no more than notation, since it is equivalent to the product of currents in \eqn{fierz}; it will be useful when summing rapidity logarithms. Typically in SCET one then factorizes $\ttwo$ into the product of two TMDPDFs, each of which is then individually renormalized (and individually ill-defined without the introduction of a rapidity regulator). Here we do not further factorize $\ttwo$, but instead treat it as a single object which has the convolution included as part of its definition.

If $q_T\sim \lqcd$, the differential rate is given by the nonperturbative matrix element of $\ttwo$ in \eqn{dsigma}.
Here we will focus on the hierarchy $q_T\gg \lqcd$, which allows us to perform an additional matching at the scale $\mu=q_T$ of $\ttwo$ onto a product of two parton distribution functions,
\begin{equation} \begin{aligned} \label{Tfact}
\ttwo(q^-,q^+,\myvec q_T)\rightarrow  &\int{dz_1\over z_1}{dz_2\over z_2} \, \ctwo\left(z_1,z_2,\myvec q_T,\mu\right) 
\\&\times O_q\left({q^-\over z_1}\right) O_{\bar q}\left({q^+\over z_2}\right)+O\left({1\over q_T^2}\right)
\end{aligned} \end{equation}
where
\begin{equation}\begin{aligned}
O_q(\ell^-)&={1\over 2\pi} \int d\xi \, e^{-i \xi \ell^-} \bar \psi_n(\nb\xi) {\nbs\over 2} W(\nb\xi, 0) \psi_n(0) \\
O_{\bar q}(\ell^+)&={1\over 2\pi} \int d\xi \, e^{-i \xi \ell^+} \bar \psi_\nb(0) {\ns\over 2} W(0,n\xi) \psi_\nb(n \xi)
\end{aligned}
\end{equation}
are the usual unpolarized lightcone distribution operators as used in Deep Inelastic Scattering \cite{Manohar:2003vb}, whose hadronic matrix elements are the parton distribution functions
\begin{equation}\begin{aligned}
\langle N(P)|O_q(\ell^-)|N(P)\rangle&=f_{q/N}\left({\ell^-\over P^-}\right)\\
\langle N(P)|O_{\bar q}(\ell^+)|N(P)\rangle&=f_{\bar q/N}\left({\ell^+\over P^+}\right).
\end{aligned}\end{equation}
The matching coefficient $\ctwo\left(\omega_1,\omega_2,\myvec q_T,\mu\right)$ may be calculated by evaluating matrix elements of both sides of \eqn{Tfact} between perturbative quark and gluon states.

\begin{figure}[h] \begin{centering}
	\includegraphics[width=0.45\textwidth]{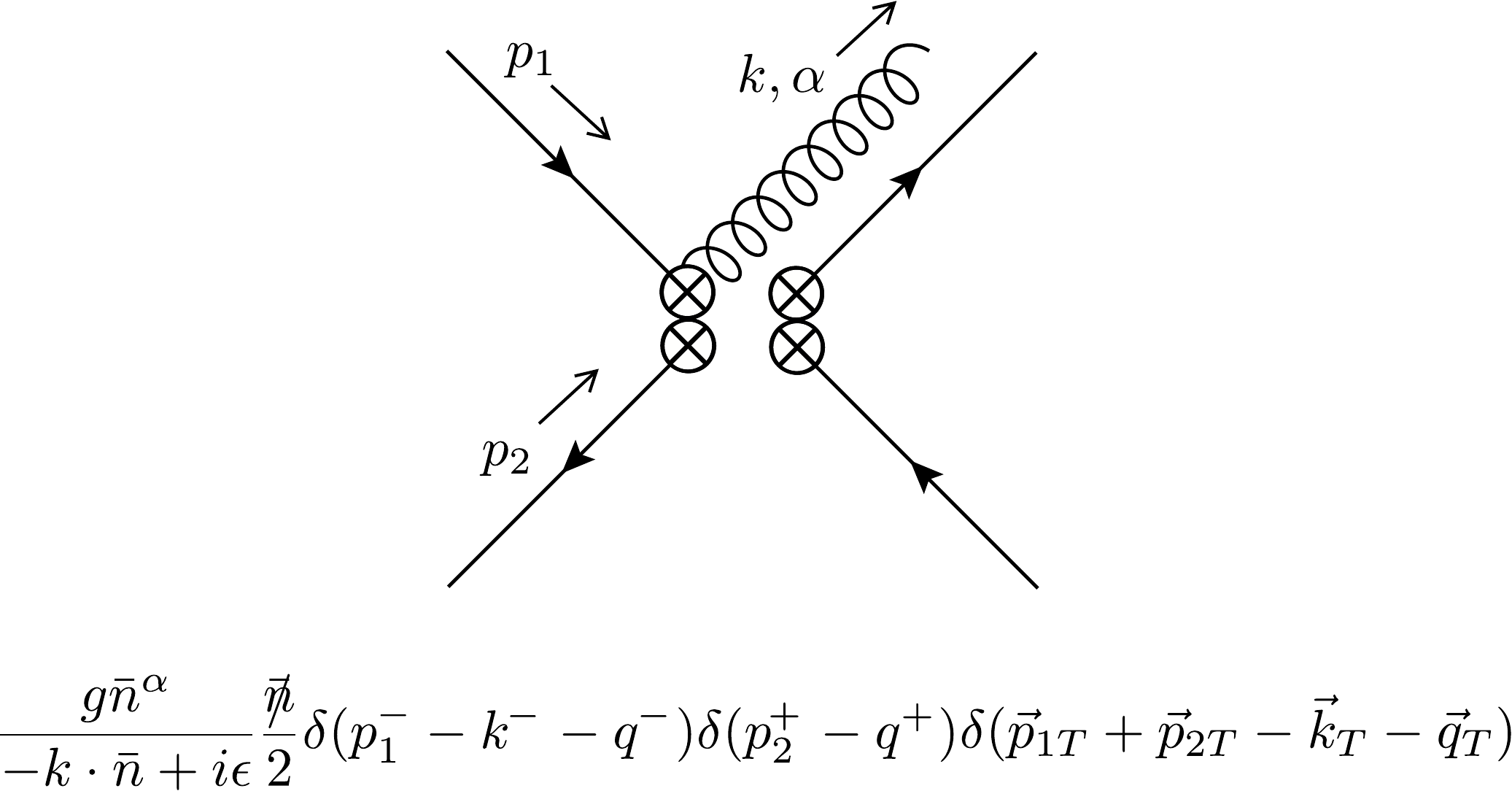}
	\caption{One-gluon Feynman rule ($n$-sector) for the left $\ttwo(q^-, q^+, \myvec q_T)$ vertex.}
	 \label{fig:wilsonfig}
\end{centering} \end{figure}

\subsection{Matrix Elements of \texorpdfstring{$\ttwo$}{T(0,0)}}

At tree level, $\ttwo$ has the spin-averaged parton-level matrix element
\begin{equation} \begin{aligned}  \label{eq:DY_SCET_tree}
\mathcal{\overline M}_0 &\equiv {1\over 4}\sum_{\rm spins}\bra{p_1 p_2} \ttwo(q^-, q^+, \myvec q_T) \ket{p_1 p_2} 
\\& =\delta( \zbar ) \delta( \zzbar ) \delta(\myvec q_T-\myvec p_{1T}-\myvec p_{2T})+O(\alpha_s)
\end{aligned} \end{equation}
where $p_1$ is the momentum of the incoming quark in the $n$-sector, $p_2$ the momentum of the antiquark in the $\bar n$-sector, and $\bar z_1\equiv 1-z_1\equiv1-q^-/p_1^-$, $\bar z_2\equiv 1-z_2\equiv 1-q^+/p_2^+$. The parton-level matrix elements of $O_q$ and $O_{\bar q}$ manifestly factorize, since the $n$ and $\bar n$ sectors are decoupled:
\begin{equation}\begin{aligned}
\langle p_1, p_2|O_q(k^-) O_{\bar q}(k^+)|p_1, p_2\rangle = & \langle p_1|O_q(k^-)|p_1\rangle 
\\&\times \langle p_2|O_{\bar q}(k^+)|p_2\rangle\\
\end{aligned}\end{equation}
and, to one loop, we have the familiar spin-averaged matrix element
\begin{equation}\begin{aligned}\label{pdfpert}
{1\over 2}\sum_{\rm spins}  \langle p_1|O_q(z p_1^-)|p_1\rangle&=\delta\left(1-z\right)-{\bar\alpha\over\epsilon}\left[{1+z^2\over 1-z}\right]_+
\end{aligned}\end{equation}
where the infrared divergent term is the usual one-loop Altarelli-Parisi splitting kernel. 

\begin{figure}[h] \begin{centering}
	\includegraphics[width=0.45\textwidth]{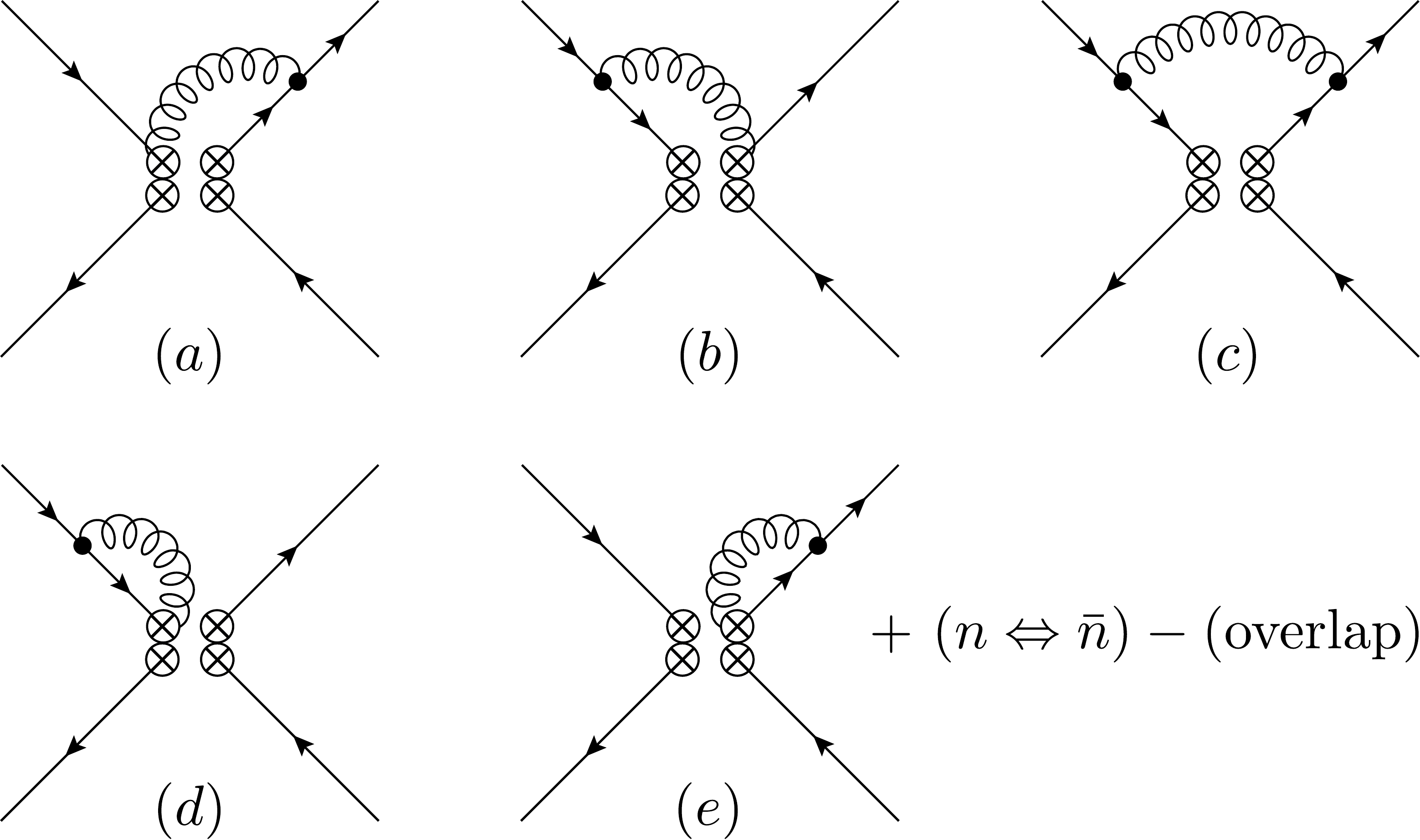}
	\caption{One-loop graphs contributing to $\langle p_1 p_2|\ttwo|p_1 p_2\rangle$.}
	 \label{fig:dygraphs}
\end{centering} \end{figure}

Expanding \eqn{eq:DY_SCET_tree} in powers of $\myvec p_{iT}/q_T$ and comparing with \eqn{pdfpert} gives the tree-level matching condition 
\begin{equation}
\ctwo^{(0)}(z_1,z_2,\myvec q_T,\mu)=\delta(1-z_1)\delta(1-z_2)\delta(\myvec q_T)
\end{equation}
where
\begin{equation} \begin{aligned}
\ctwo(z_1,z_2,\myvec q_T,\mu)\equiv &\ctwo^{(0)}(z_1,z_2,\myvec q_T,\mu) 
\\&+\bar\alpha \ctwo^{(1)}(z_1,z_2,\myvec q_T,\mu)+O(\alpha_s^2).
\end{aligned}\end{equation}
To calculate the matching at one loop, we need the matrix element of $\ttwo$ between quark states given by the diagrams in Fig. \ref{fig:dygraphs}, along with the analogous graphs with $\bar n$-sector gluons coupling to the antiquark lines. The overlap graphs are obtained from the $n$- (or equivalently $\nbar$) sector graphs by replacing the quark propagators with the corresponding lightlike Wilson lines. Since we are working at leading order in $1/q$ we may set the external transverse momenta $\myvec p_{1T}$ and $\myvec p_{2T}$ to zero.

Away from $\bar z_1=\bar z_2=0$, only graphs (a-c) and the corresponding $\nb$-sector graphs contribute, and the individual graphs are well-defined. The only component of the gluon loop momentum $k^\mu$ not fixed by energy-momentum conservation is $k^-$, so the loop integral is easily done using contour integration\footnote{Note that these are not cut graphs, but the poles from the light quark propagators are on the opposite side of the real axis from the pole from the gluon propagator, so the $k^-$ integral picks out only the pole at $k^2=0$.}. We use the distributional relation in $d$ dimensions,
\begin{equation}\label{lndef}
{1\over \myvec{q}_T^{\,2} }=-{S_{2-2\epsilon}\over 2\epsilon}\mu^{-2\epsilon}\delta(\myvec q_T)+\plusmu{1\over \myvec{q}_T^{\,2}}
\end{equation}
where $S_d=2\pi^{d/2}/\Gamma(d/2)$, and $\plusmu{1\over \myvec{q}_T^{\,2}}$ 
denotes a $(d-2)$-dimensional plus distribution, defined as
\begin{equation}
\begin{aligned}
\plusmu{g(\myvec p_T)}&=g(\myvec p_T),\ |\myvec p_T|>0, \ \ \mbox{and}\\
\int_{|\myvec p_T|\leq\mu} d^{d-2} \myvec p_T\plusmu {g(\myvec p_T)}&=0.
\end{aligned}
\end{equation}
(Note that \eqn{lndef} is independent of $\mu$.) After renormalization we take the limit $d-2\rightarrow 2$, which recovers the 2-dimensional plus distribution definition of \cite{Ebert:2016gcn}.
This gives for the spin-averaged $n$-sector graphs 
\begin{equation} \begin{aligned}  \label{eq:DYambig_n}
\mathcal{\overline M}_{1n}(\bar z_1\neq 0)&=  {\alphabar\over\pi} f_\epsilon\left(-{S_{2-2\epsilon}\over 2\epsilon}\mu^{-2\epsilon}\delta(\myvec q_T)+\plusmu{1\over \myvec{q}_T^{\,2}}\right) 
\\&\times \delta(\zzbar) \frac{ 2-2\zbar+\zbar^2(1-\epsilon)}{\zbar},
\end{aligned} \end{equation}
where $f_\epsilon=\pi^\epsilon \mu^{2\epsilon} e^{\epsilon\gamma_E}$, while the $\nb$ graphs yields the same under the switch $\zzbar\leftrightarrow \zbar$.  We can therefore write the spin-averaged one-loop matrix element of $\ttwo$ as the sum of contributions away from $\zbar=\zzbar=0$ and some unknown contribution at this point,
\begin{equation} \begin{aligned} \label{t22oneloop}
\mathcal{\overline M}_1=  & {\alphabar\over\pi} f_\epsilon
\left(
    -{S_{2-2\epsilon}\over 2\epsilon}\mu^{-2\epsilon}\delta(\myvec q_T)
    +\plusmu{1\over \myvec{q}_T^{\,2}}
\right) 
\\&\times
\bigg(
    A \delta(\zbar)\delta(\zzbar)
    +\delta(\zzbar) \left[\frac{ 2-2\zbar+\zbar^2(1-\epsilon)}{\zbar}\right]_+ 
    \\&\qquad
    +\delta(\zbar) \left[\frac{ 2-2\zzbar+\zzbar^2(1-\epsilon)}{\zzbar}\right]_+
\bigg)
\end{aligned} \end{equation}
then calculate the contributions to the constant $A$  from graphs (a-c) by integrating the individual graphs with respect to $\zbar$ and $\zzbar$ before doing the loop integrals. The gluon momentum $k^-$ is then no longer fixed by the delta functions, and
just as in the previous section, the integrals defining $A$ in \eqn{t22oneloop} contain rapidity divergences and are not individually well-defined. 

As in the case of the massive Sudakov form factor, it is instructive to follow the na\"\i ve scheme of performing the integrals for all graphs except for the $k^-$ integral. In this case, the contributions to $A$ from graphs (a-c) and their counterparts in the $\nbar$ and overlap sectors are
\begin{equation} \begin{aligned}  \label{eq:DYambig_INT}
A_n&=   \int_0^{p_1^-} \frac{d k^-}{k^-} \left[ 
2 - 2\frac{k^-}{ p_1^- }  + \left( \frac{k^-}{p_1^-}\right)^2 (1-\epsilon) \right] \\
A_\nb&=  
\int_\frac{q_T^2}{p_2^+}^{\infty} \frac{d k^-}{k^-} \left[ 2  -2 \left(\frac{q_T^2 }{ k^- p_2^+}\right) + (1-\epsilon)\left(\frac{ q_T^2 }{k^- p_2^+} \right)^2  \right] \\
A_o&= 2\int_0^\infty \frac{d k^-}{k^-} 
\end{aligned} \end{equation}
where $\shat\equiv p_1^- p_2^+=q^2/(z_1 z_2)$. 
Proceeding as in the previous section, we can rescale the integration variable 
$k^- \to \zeta^2 k^-\! / \shat$
in the $A_\nb$ integral in \eqn{eq:DYambig_INT} to obtain
the scheme-dependent sum
\begin{equation}\label{azetaR}
A(\zeta)\equiv A_n+A_\nb-A_o=2\log{\zeta^2\over q_T^2}-3-\epsilon.
\end{equation}
Again, we see that the SCET calculation has no dynamical dependence on the hard scale $\hat{s}$; rather, it arises as one possible choice of scheme needed to evaluate the sum of divergent integrals. 

As in the previous section, we can make the individual graphs well-defined with an appropriate regulator. Modifying the position-space Wilson lines to include the $\delta$-regulator, the $n$-sector contributes
\begin{equation} \begin{aligned}  \label{eq:DYn_z}
\overline{\cal M}_{1n}^\delta = &{\alphabar\over\pi} f_\epsilon\left(-{S_{2-2\epsilon}\over 2\epsilon}\mu^{-2\epsilon}\delta(\myvec q_T)+\plusmu{1\over \myvec{q}_T^{\,2}}\right)
\\&\times
\delta(\zzbar)
\left( \frac{ 2(1-\zbar) }{\zbar + \frac{ \delta_n}{p_1^-} }  + \zbar(1-\epsilon)  \right).
\end{aligned} \end{equation}

\hspace{-0.11in}Converting to distribution form using 

\begin{equation} \begin{aligned}  \label{eq:delta_plus_identity}
\frac{1}{\bar{z} + \delta }  =&- \delta\left(\bar{z}\right) \log \delta  + \left[{1\over \bar z}\right]_+ +O(\delta)
 \end{aligned} \end{equation}
gives
\begin{equation} \begin{aligned}  \label{eq:DYn_final}
\overline{\cal M}_{1n}^\delta& =  \alphabar \delta(\zzbar)
\Bigg\{
    \zbar\delta(\myvec q_T)
    +
    \left(  \frac{ \delta(\myvec q_T)}{\epsilon} -  {1\over\pi}\plusmu{1\over \myvec{q}_T^{\,2}} \right)
    \\&\qquad\times     
    \left( 
        \delta(\zbar) 
        \left[ 
            \frac{3}{2} -2\log \frac{p_1^-}{\delta_n}
        \right] 
        - \plusfn{1+z_1^2\over \zbar}  
    \right)
\Bigg\}
\end{aligned} \end{equation}

The $\nb$-sector graphs are the same as above but with $p_1^- \rightarrow p_2^+$ and $z_1 \leftrightarrow z_2$, and the calculation of the overlap graphs gives
\begin{equation} \begin{aligned}  \label{eq:DYover_dkm}
\overline{\cal M}_{1o}^\delta
= 2{\alphabar\over\pi} f_\epsilon\delta(\zbar) \delta(\zzbar)  \frac{1}{q_T^2} \log\frac{q_T^2}{\delta_o^2} .
\end{aligned} \end{equation}
In $d$ dimensions,
\begin{equation}
{\log{q_T^2\over \mu^2}\over q_T^2}= -{S_{2-2\epsilon}\over2\epsilon^2}\mu^{-2\epsilon}\delta(\myvec q_T)+\plusmu{\log{\myvec{q}_T^{\,2}\over \mu^2}\over \myvec{q}_T^{\,2}}\end{equation}
\ \\ 
and so 
\begin{equation} \begin{aligned}  \label{eq:DYover_final}
\overline{\cal M}_{1o}^\delta &=  
\alphabar\delta(\zbar) \delta(\zzbar) 
\Bigg[ 
    \delta(\myvec q_T)\bigg( -\frac{2}{\epsilon^2} + \frac{2\log\frac{\delta_o^2}{\mu^2}}{\epsilon} +{\pi^2\over 6}\bigg) 
    \\& -{2\over\pi}\log\frac{\delta_o^2}{\mu^2} \plusmu{1\over \myvec{q}_T^{\,2}} 
    + {2\over\pi} 
    \bigg[ 
        \frac{   \log \frac{\myvec{q}_T^{\,2}}{\mu^2}  }{\myvec{q}_T^{\,2}} 
    \bigg]_+^{\mu^2}
\Bigg]
\end{aligned} \end{equation}

Combining all the above yields the net contribution to matrix elements of $\ttwo$ from single-gluon emissions into the final state
\begin{widetext}
\begin{equation} \begin{aligned}  \label{eq:DY_delta_final}
\overline{\cal M}_{1g}^\delta =&   \alphabar \left(
\vphantom{\plusmu{\log{\myvec{q}_T^{\,2}\over\nu^2}\over \myvec{q}_T^{\,2}}}
 \delta( \zbar ) \delta( \zzbar ) \delta(\myvec q_T) \left( 
\frac{2}{\epsilon^2} + \frac{3-2\log \frac{\nu^2}{\mu^2}}{\epsilon}-{\pi^2\over 6} \right)\right.  
\\ &
- \frac{ \delta(\myvec q_T) }{\epsilon} \left( \delta(\zbar)\plusfn{\frac{1+z_2^2}{\zzbar}} + \delta(\zzbar) \plusfn{\frac{1+z_1^2}{\zbar}}\right) \\
& + \left( (1+z_2^2) \delta( \zbar ) \plusfn{\frac{1}{\zzbar}} + (1+z_1^2) \delta( \zzbar ) \plusfn{\frac{1}{\zbar}} \right)  {1\over\pi}\plusmu{{1\over \myvec{q}_T^{\,2}}}\\
&+ \delta(\myvec q_T)\left( \zzbar \delta( \zbar ) + \zbar \delta(\zzbar ) \right)  \left.- 2 \delta( \zbar )\delta(\zzbar ){1\over\pi}\plusmu{\log{\myvec{q}_T^{\,2}\over\nu^2}\over \myvec{q}_T^{\,2}}
\right)
\end{aligned} \end{equation}
\end{widetext}
where, as with the form factor calculation, we have taken the limit $\delta_{n,\nb,o}\to 0$, again holding the ratio $\delta_n \delta_\nb/\delta_o^2 \equiv q^2/\nu^2$  fixed, which defines the scheme for regulating the rapidity divergences. The virtual graphs are scaleless, so do not contribute in this scheme.

The divergences in the first line in \eqn{eq:DY_delta_final} are cancelled by the counterterm for $O_2$ if $\nu^2=q^2$, which imposes the scheme $\nu=q$ when matching from QCD onto  SCET. The divergence in the second line is equal to the infrared divergence in the matrix elements of $O_q$ and $O_{\bar q}$ in \eqn{pdfpert} and cancels in the matching conditions at $\mu=q_T$, leaving the one-loop result
\begin{widetext}
\begin{equation}\begin{aligned}
\ctwo^{(1)}(z_1, z_2, \myvec q_T,\mu)=& \left( \left(1+z_2^2\right) \delta(\zbar) \plusfn{\frac{1}{\zzbar}} + \left(1+z_1^2\right)\delta(\zzbar )\plusfn{\frac{1}{\zbar}} \right) {1\over\pi}\plusmu{\frac{1}{ \myvec{q}_T^{\,2}}} \\
&+  2\delta(\zbar)\delta(\zzbar)  \log{q^2 \over \mu^2 } \, {1\over\pi}\!\plusmu{  \frac{1}{\myvec{q}_T^{\,2}}  }  - 2 \delta(\zbar )\delta(\zzbar) {1\over \pi}\! \plusmu{\frac{\log\frac{\myvec{q}_T^{\,2}}{\mu^2}}{\myvec{q}_T^{\,2}}}\\
&+ \delta(\myvec q_T)\left( \zzbar \delta(\zbar) + \zbar \delta(\zzbar) - {\pi^2\over 6} \delta(\zbar) \delta(\zzbar)\right)  
\end{aligned} \end{equation}
\end{widetext}
where $\bar\omega_{1,2}\equiv1-\omega_{1,2}$.
The rapidity logarithm depends on the ultraviolet scale $q^2$ only because of the choice of scheme parameter $\nu^2=q^2$. As in the previous section, we may resum the leading order rapidity logarithms by running the scheme parameter $\nu$ from $\nu_H^2=q^2$ to $\nu_S^2=q_T^2$, as will be discussed in the next section.

%
%
\subsection{Resummation}
%
%

The rapidity renormalization group equations for $\ttwo$, which resum the large logarithms of $q^2/q_T^2$ in $\ctwo$, arise from the independence of $\ctwo$ on the scheme parameter $\nu$.  Since matrix elements of $\ttwo$ have no dynamical dependence on $q^2$, we can write, proceeding analogously to \eqn{o2match},
\begin{equation} \begin{aligned} \label{t22fact}
&\ttwo(q^+, q^-, \myvec q_T,\nu^2=q^2)|_{\mu=q_T}
\\&\quad 
=\int{d\omega_1\over\omega_1}  {d\omega_2\over\omega_2} d^2\myvec{p}_T \ 
\dtwo\!\left(\omega_1, \omega_2, \myvec{p}_T,{q^2\over\nu^2}\right) 
\\ &\qquad\times
\ttwo\left({q^+\over\omega_1}, {q^-\over\omega_2},  \myvec{q}_T-\myvec{p}_T,\nu^2 \right)
\end{aligned} \end{equation} 
%
and, from \eqn{eq:DY_delta_final}, at one loop
\begin{equation} \begin{aligned}  \label{eq:DY_SCET_OET_matching}
\dtwo(\omega_1,\omega_2,\myvec{p}_T,\nu)=& \delta(\omegaoneb)\delta(\omegatwob)
\Bigg( 
    \delta(\myvec{p}_T)
    \\&
    +2\alphabar  \log\frac{q^2}{\nu^2}\, \frac{1}{\pi} \left[ \frac{1 }{\myvec{p}_T^{\,2}}\right]_+^{\mu} 
\Bigg)+ \dots. \end{aligned} \end{equation}

\eqn{t22fact} plays the role of a SCET factorization theorem in this analysis, although here it just reflects the fact that $\ttwo$ is the only operator at leading order contributing to the cross section, so the cross section must be expressible as a linear combination of $\ttwo$'s (with different arguments).
Since \eqn{t22fact} is independent of $\nu$, we perform the standard manipulations and find
\begin{equation} \begin{aligned}  \label{eq:DY_simplified_running}
\frac{d}{d\log\nu} \dtwo(\myvec{p}_T,\nu) &= \int d^2 \myvec{k}_T \!  
\left( -4\alphabar \frac{1}{\pi} \plusmu{\frac{1}{(\myvec{p}_T-\myvec{k}_T)^2}}\right)  
\\&\qquad\times
\dtwo(\myvec{k}_T,\nu) \\
&\equiv \gtwo(\myvec{p}_T )\otimes \dtwo(\myvec{p}_T,\nu) \\
\end{aligned} \end{equation}
where the two dimensional convolution is defined as
\begin{equation}
f(\myvec{p}_T)\otimes g(\myvec p_T, ..)\equiv \int d^2\myvec k_T \, f(\myvec p_T-\myvec k_T, ...)g(\myvec k_T, ...) .
\end{equation}
We have made use of the fact that at one loop the anomalous dimension is diagonal in each of the $\omega_i$,\ and defined
$\dtwo(\omega_1,\omega_2,\myvec p_T,\nu)\equiv  \delta(\omegaoneb)\delta(\omegatwob)\dtwo(\myvec{p}_T,\nu)$. 
$\dtwo$ therefore obeys the same form of rapidity renormalization group equation as a beam function in the usual SCET formalism.  The all-orders solution for $V$ and the complications associated with evaluating distributions at canonical scales has been discussed in detail in \cite{Ebert:2016gcn}.

This gives the resummed formula for the low-scale matching condition 
\begin{equation} \begin{aligned}
\ctwo(z_1,z_2,\myvec q_T,\mu_S)&= V(\myvec q_T,\mu_S, \nu_S)
\\ & \quad \otimes 
\ctwo(z_1,z_2,\myvec q_T, \mu_S,\nu_S)
\end{aligned} \end{equation}
and the final factorized and resummed expression for the DY cross section reads
\begin{widetext}
\begin{equation}\begin{aligned}\label{dsigmaressumed}
{d^4\sigma\over dq^+ dq^- d^2\myvec q_T}={4\pi\alpha^2\over 3N_c q^2 s}
|U_2(\mu_S,\mu_H)C_2(\mu_H)&|^2
\int {dz_1\over z_1}{dz_2\over z_2} V(\myvec q_T,\mu_S,\nu_S) \otimes \ctwo(z_1,z_2,\myvec q_T, \mu_S,\nu_S) f_{q}\left({\xi_1\over z_1}\right) f_{\bar q}\left({\xi_2\over z_2}\right)
\end{aligned}
\end{equation}
\end{widetext}
where $\xi_1=q^-/P_1^-$, $\xi_2=q^+/P_2^+$, $\mu_H=\nu_H=q$, $\mu_S=\nu_S=q_T$, and $U_2$ is defined in \eqn{eq:U2evolution}.

\section{Conclusions}\label{sec:Conclusion}

In this paper we have demonstrated how to apply SCET to processes involving rapidity divergences without explicitly separating the low energy degrees of freedom into separate modes. We have shown that the anomalous appearance of 
the hard scale $Q$ in the effective theory below $Q$ in \scettwo-type problems arises from a scheme dependence in the effective theory. This scheme dependence is common for both \scetone\ and \scettwo\  processes, with the only distinction between these types of processes being whether the matching coefficient onto the soft theory exhibits a large logarithmic enhancement (\scettwo) or not (\scetone); the intermediate effective theory is the same until we reach the matching scale at which the process dependence arises. The free scheme parameter can be exploited to derive evolution equations for matching coefficients, yielding a method for the summation
of the large rapidity logarithms which appear in the soft matching coefficients of \scettwo\ processes. 

The factorizations and resummations presented in this paper are well known in the standard SCET formalism, and we reproduce the results here.  However, reducing the number of distinct fields in the theory simplifies the structure of the theory and significantly reduces the number of Feynman diagrams and operators required for a given calculation. In particular, we expect the calculation of power corrections in SCET, which have been recently of much interest \cite{Beneke:2018gvs,Ebert:2018gsn,Moult:2018jjd,Moult:2019vou}, to be significantly simplified.  The matching and anomalous dimensions of power-suppressed contributions to the dijet current were calculated in this formalism in \cite{Goerke:2017lei}; Fierz-rearranged  products $T_{(i,j)}$ of these subleading operators, analogous to $\ttwo$, may be constructed and their rapidity logarithms resummed by exploiting the scheme dependence of the rapidity regulator. However, as pointed out in \cite{Ebert:2018gsn}, at subleading orders in $1/Q$ the $\delta$-regulator is not sufficient to regulate all the rapidity divergences and another regulator, such as the pure rapidity regulator presented in that reference, is required.  Work on this subject is in progress.

\begin{acknowledgments}

We thank Jyotirmoy Roy for useful discussions of this and related work. This work was supported in part by the Natural Sciences and Engineering Research Council of Canada. 
\end{acknowledgments}


\bibliography{scet2_PRD_rev}

\end{document}